\documentclass[twocolumn,prb,showpacs,preprintnumbers,superscriptaddress,amsmath,amssymb,citeautoscript]{revtex4-1}
\usepackage{graphicx}
\usepackage{epstopdf}
\usepackage{dcolumn}
\usepackage{color}
\usepackage{amsmath,amssymb}
\usepackage[english]{babel}
\usepackage{bm}
\usepackage[hyperindex]{hyperref}
\hypersetup{colorlinks,citecolor=blue, filecolor=blue,linkcolor=blue,urlcolor=blue}

\usepackage{multirow}
\usepackage{rotating}
\usepackage{array}

\newcommand{\bpm}{\begin{pmatrix}}
\newcommand{\epm}{\end{pmatrix}}

\newcommand{\be}{\begin{equation}}
\newcommand{\ee}{\end{equation}}
\newcommand{\beq}{\begin{eqnarray}}
\newcommand{\eeq}{\end{eqnarray}}

\DeclareMathOperator{\sgn}{sgn}

\DeclareMathOperator{\im}{Im}

\DeclareMathOperator{\tr}{tr}
\DeclareMathOperator{\T}{T}

\begin{document}

\title{Modeling long imperfect SNS junctions and Andreev bound states using two impurities and the \boldmath{$T$}-matrix formalism}
\author{Sarah Pinon}
\affiliation{Institut de Physique Th\'eorique, Universit\'e Paris Saclay, CEA
CNRS, Orme des Merisiers, 91190 Gif-sur-Yvette Cedex, France}
\author{Vardan Kaladzhyan}
\email{vardan.kaladzhyan@phystech.edu}
\affiliation{Department of Physics, KTH Royal Institute of Technology, Stockholm, SE-106 91 Sweden}
\author{Cristina Bena}
\affiliation{Institut de Physique Th\'eorique, Universit\'e Paris Saclay, CEA
CNRS, Orme des Merisiers, 91190 Gif-sur-Yvette Cedex, France}

\date{\today}

\begin{abstract}
We provide a new analytical tool to calculate the energies of Andreev bound states (ABS) in long imperfect SNS junctions, at present these can only be described by numerical tools. We model an NS junction as a delta-function ``Andreev'' impurity, i.e., a localized potential which scatters an electron into a hole with opposite spin. We show using the scattering matrix formalism that, quite surprisingly,  an ``Andreev'' impurity is equivalent to an NS junction characterized by both Andreev reflection and a finite amount of normal scattering. The ABS energies are then calculated using the $T$-matrix formalism applied to a system with two Andreev impurities. Our results lie between those for a perfect long SNS junction limit described by the Andreev approximation (ABS energies depend linearly on the phase and are independent of the chemical potential) and the particle-in-the-box limit (bound state energies are independent of the phase and have a linear dependence on the chemical potential). Moreover, we recover a closed-form expression for the ABS energies by expanding around the particle-in-the-box limit.
\end{abstract}

\maketitle

\section{Introduction}

The formation of Andreev bound states (ABS) \cite{Andreev1964,Kulik1970,Beenakker1991,Fukuyama1992,Beenakker1992} in long SNS junctions\cite{Griffin1971,Josephson1974,Blonder1982,deGennes1999} has been approached by analytical tools such as the Bogoliubov--de Gennes equations \cite{Furusaki1991,Hurd1994,Richter2002}, the Andreev approximation\cite{Affleck2000,Nazarov2009,Beenakker2013}, as well as various other approaches \cite{MartinRodero1994,LevyYeyati1995,Fazio1995,Cuevas1996,Caux2002,Bauer2007,Perfetto2009,Meng2009}.
However, taking into account a finite amount of backscattering at the leads in such junctions has so far only been possible using numerical tools (see, e.g., Ref.~[\onlinecite{Bena2012}]). For instance there is no analytical description of the bell-shaped dependence of the ABS in long junctions on gate-voltage, neither of their non-linear dependence on the phase\cite{Bena2012,Pillet2010,Dirks2011}. 

Here we provide the first fully analytical tool that can take into account the normal scattering at the leads in long junctions (in the limit of energies much smaller than the superconducting gap). 
When there is no normal scattering at the leads, the physics of the ABS in this limit is described by the Andreev approximation  \cite{Affleck2000,Nazarov2009,Beenakker2013}, yielding a linear dependence of the ABS energies on the phase; in this limit their energies are independent of the chemical potential. In the opposite limit, when the normal scattering at the leads is very large, we recover the particle-in-the-box limit, i.e., the normal region is completely isolated and the energies of the bound states correspond to the quantized levels of a finite-size system; the energies of these states depend linearly on the gate voltage and are independent of the SC phase difference. For the intermediate regime, in which the normal scattering is finite, we show that the energies of the ABS lie between the two limits and we obtain a closed-form expression for these energies by expanding around the particle-in-the box limit. This expression allows to extract information  for example about the periodicity and the amplitude of the ABS oscillations with the chemical potential, the amplitude of the ABS oscillations with the phase, and the width of the ABS levels as a function of the scattering strength; such information will be useful in order to analyze experimental observations of the ABS in long imperfect SNS junctions, once such observations become available.


We will first show that we can model an imperfect NS junction by considering an ``Andreev'' type impurity: a delta-function localized potential that scatters an electron into a hole with opposite spin. This equivalence is demonstrated using the scattering formalism: we find that there are regimes of parameters in which the values of the reflection and Andreev reflection coefficients in the NS junction can be the same as those generated by an ``Andreev'' impurity. In this regime an ``Andreev'' impurity can thus be a good model for an NS junction characterized by both Andreev reflection and a finite amount of normal scattering. Our formalism cannot describe the ideal NS junctions, a finite amount of normal scattering at the leads is automatically included.

\begin{figure}[ht!]
\centering
\includegraphics[width=0.92\columnwidth]{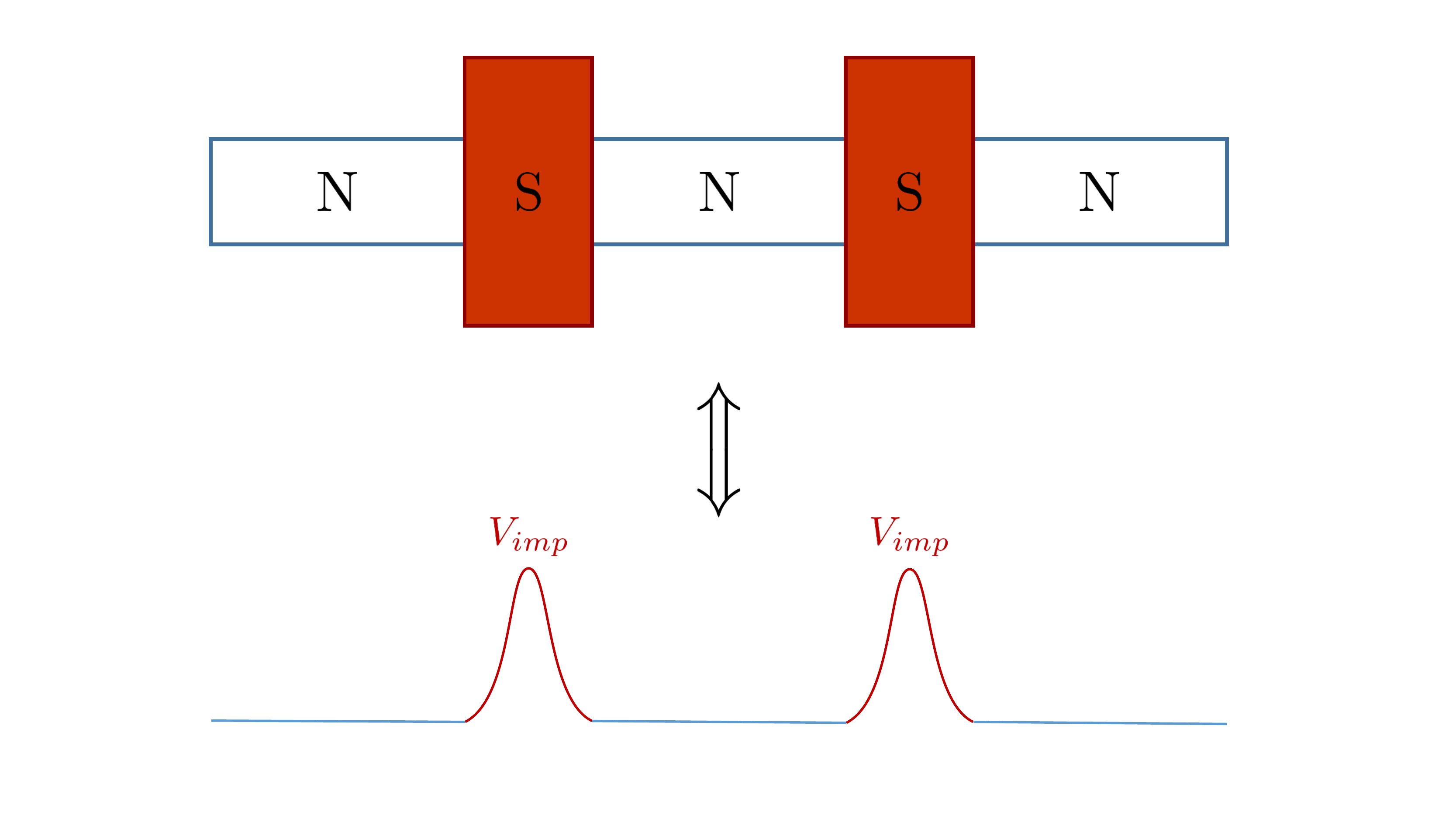}
\caption{Schematics of the equivalence between `Andreev' impurities and NS junctions.}
\label{fig:SNSasAndreevImpurities}
\end{figure}

However, the point of this work is not to model a sole NS junction, for which many alternative analytical works exist, but to model an imperfect SNS junction, or equivalently, two imperfect SN junctions situated at a finite and large distance from each other, for which it is to our knowledge impossible to obtain fully analytically the form of the local density of states or the energy dependence of the ABS. We thus model the SNS junction by considering two ``Andreev'' impurities situated at a given distance, as illustrated in Fig.~\ref{fig:SNSasAndreevImpurities}. We use the $T$-matrix formalism which provides an exact solution for the two-impurity problem. The resulting Green's function in the region between the two impurities corresponds to that of a normal region in an SNS junction, and thus it allows us to describe the formation of ABS. We calculate the resulting dependence of the ABS energies on various parameters such as the gate voltage, the phase difference between the two SCs and the normal scattering at the leads, and we show that it is consistent with previous findings (see, for instance, the results obtained by numerical diagonalization in Ref.~[\onlinecite{Bena2012}]).

In Sec.~II we provide the equivalence conditions between the NS junction and the ``Andreev'' impurity. In Sec.~III we give the $T$-matrix formalism necessary to compute the Green's function and subsequently the local density of states (LDOS) in the presence of two impurities. We present our results in Sec.~IV, leaving the conclusions to Sec.~V.

\section{Scattering formalism and conditions of equivalence between the NS junction and the ``Andreev'' impurity}

\begin{figure}[t!]
\centering
\includegraphics[width=8cm,trim={2cm 0cm 2cm 0cm},clip]{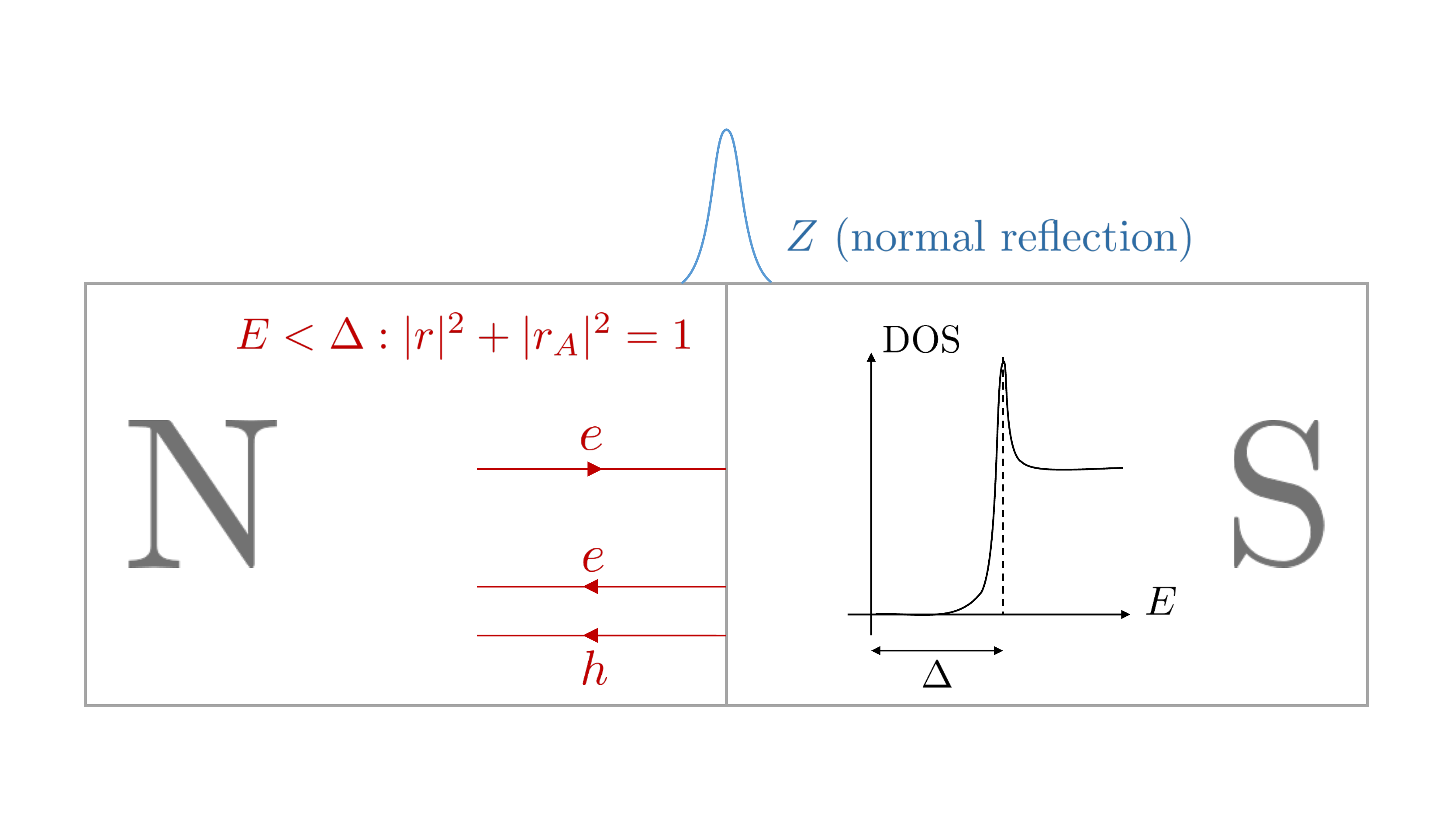}
\caption{Schematics of an NS junction where we consider $E$ to be inferior to the SC gap, i.e., the transmission coefficients $t,t_A=0$.}
\label{fig:NSjunctionSchematic}
\end{figure}

For an NS junction the BTK theory\cite{Blonder1982} indicates that the reflection and Andreev reflection coefficients, resulting from injecting an electron from the normal side of the junction (see Fig. \ref{fig:NSjunctionSchematic} for the schematics of the junction), are given by

\begin{align}
|r_A|^2 &= \frac{\Delta^2}{E^2+(\Delta^2-E^2)(1+2 Z^2)^2}, \\
|r|^2 &= 1-|r_A|^2,
\end{align}
where $\Delta$ is the SC gap, $Z$ is the dimensionless barrier strength introducing a finite amount of normal reflection at the interface. We will focus only on energies much smaller than the gap, thus we have
\begin{align}
\label{eq:ZparameterNSjunction} 
|r_A|^2 &= \frac{1}{(1+2 Z^2)^2},\\
|r|^2 &= 1-|r_A|^2.
\end{align}
For $Z=0$ we have a perfect Andreev junction, i.e., $r_A=1$, while for $Z\gg1$ we have a bad junction with a lot of normal reflection.

\begin{figure}[ht!]
\centering
\includegraphics[width=7cm,trim={6cm 4cm 6cm 2cm},clip]{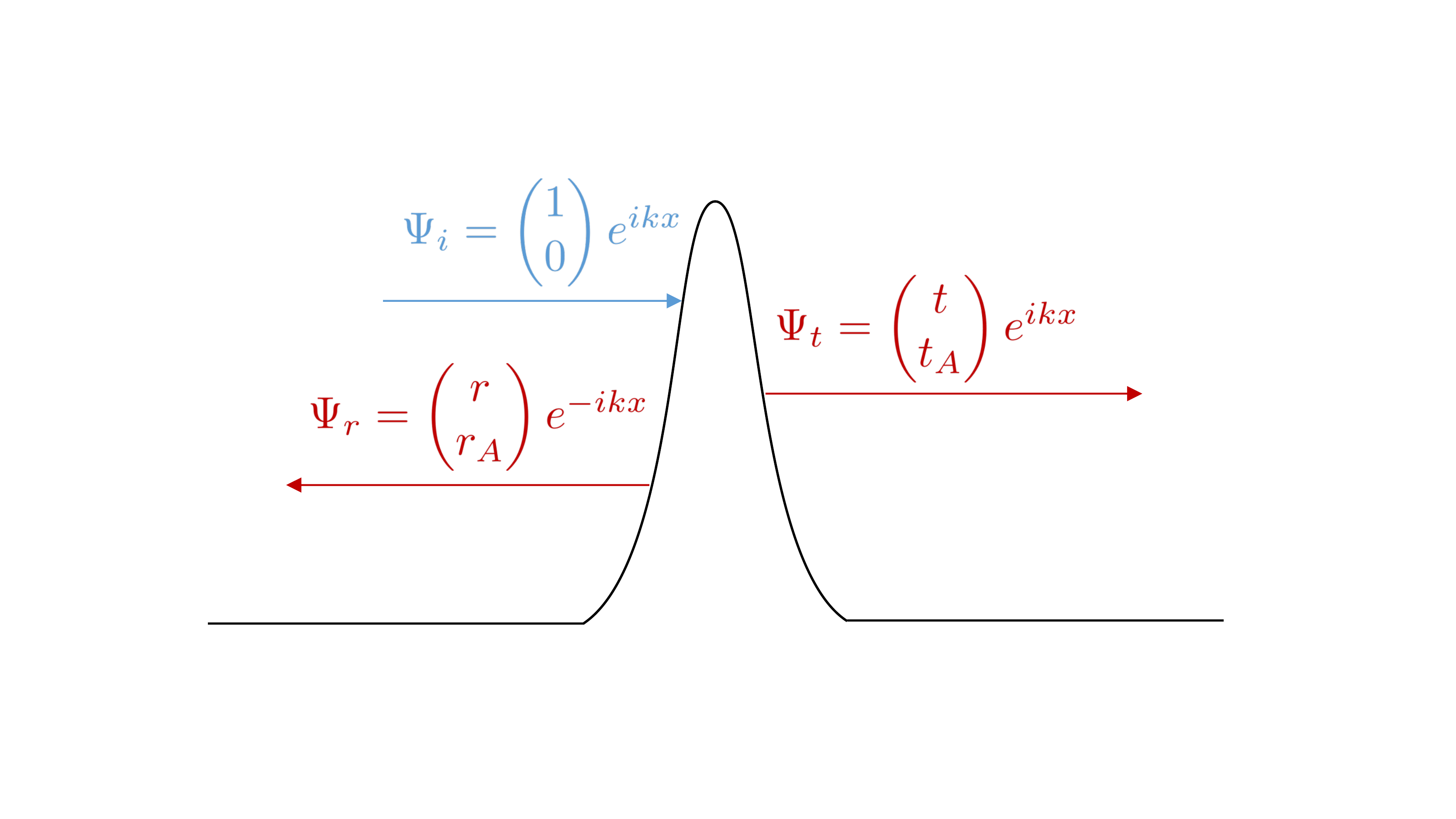}
\caption{Schematics of an ``Andreev'' impurity with the incoming, reflected and transmitted plane waves.}
\label{fig:AndreevImpurity}
\end{figure}

In order to verify the equivalence between an ``Andreev'' impurity and an NS junction we will calculate the reflection and transmission coefficients, as well as the Andreev reflection and transmission for the ``Andreev'' impurity, and check in which regime they correspond to the traditional BTK values mentioned above. 
Thus, for a delta-function Andreev impurity (see Fig. \ref{fig:AndreevImpurity}) we write down the Schr\"odinger equation
\begin{equation}
-\frac{\hbar^2}{2 m} \frac{d^2 \psi}{d x^2}(x)+V_{\mathrm{imp}}(x) \psi(x)=E \psi(x),
\label{eq:SchroedingerEq}
\end{equation}
where $\psi(x)=\left( \psi_e(x), \psi_h(x)\right)^{\T}$ is a two-component wave function with the upper and lower components standing for the electron and hole wave functions, respectively. We assume a quadratic dispersion with $m$ denoting the quasiparticle mass. Hence, the impurity potential is a $2 \times 2$ matrix in the electron-hole space, and can be written as 
\begin{align}
V_{\mathrm{imp}}(x)= V  \delta(x) \equiv 
	\bpm
	V_N & V_A \\
	V^*_A & -V_N
	\epm
	\delta(x).
\label{eq:AndreevImpPotential}
\end{align}
In the most general case a normal reflection component $V_N \neq 0$ should be present, however, we have shown that it hinders the equivalence, and therefore, hereinafter we will set $V_N=0$. Note that this does not imply that in the equivalent SN junction there is no normal scattering, in fact $V_A$ alone is generating both an Andreev and a normal component in the equivalence.

For $x \neq 0$ the solution of Eq.~(\ref{eq:SchroedingerEq}) is just a linear combination of two-component vectors multiplied by right-moving and left-moving plane waves, $e^{i k x}$ and $e ^{-i k x}$, correspondingly, where the wavevector $k \equiv \sqrt{2m  E}/\hbar$. The most general solution in the presence of the impurity is
\begin{align}
\psi_L(x)=A_r e^{i k x}+A_l e^{-i k x},\; x<0, \\ 
\psi_R(x)=B_r e^{i k x}+B_l e^{-i k x},\; x>0. 
\end{align}
We focus on the simple case of an injected electron incoming on the barrier from the left, thus $A_r=(1,0)^{\T}$, and no injected electron or hole from the right, i.e.,  $B_l=(0,0)^{\T}$. Hence $A_l=(r, r_A)^{\T}$, where $r$ is the regular reflection coefficient, and $r_A$ is the Andreev reflection coefficient. Moreover, $B_r=(t,t_A)^{\T}$ with $t$ being regular transmission and $t_A$ Andreev transmission.

By assuming that the wave function $\psi(x) = \psi_L(x)\Theta(-x)+\psi_R(x)\Theta(x)$ is continuous at $x=0$, we obtain $1+r=t$, $r_A=t_A$. We then write down the continuity equation for the derivative of the wave function. The delta-function potential gives rise to a discontinuity at the origin, proportional to the value of the impurity potential, in occurence here:
\begin{align}
-\frac{\hbar^2}{2m}[\psi'_R(0)-\psi'_L(0)]+V \psi(0)=0,
\end{align}
which yields
\begin{align}
-\frac{\hbar^2}{2m} i k (-A_r+A_l+B_r-B_l)+V (A_r+A_l) = 0.
\end{align}
We note that the solution to this problem is energy-dependent, but since we are interested in energies very close to the Fermi level, we can take $k$ to be constant $k=k_F$, and we denote $\alpha \equiv \frac{\hbar^2k_F^2}{m}$. The above equations yield 
\begin{align}
r &= -\frac{|V_A|^2}{\alpha^2+|V_A|^2}, \\
\label{eq:rA} r_A &= -i \frac{\alpha V_A^*}{\alpha^2+|V_A|^2}.
\end{align}

\begin{figure}[ht!]
\centering
\includegraphics[width=0.7\columnwidth]{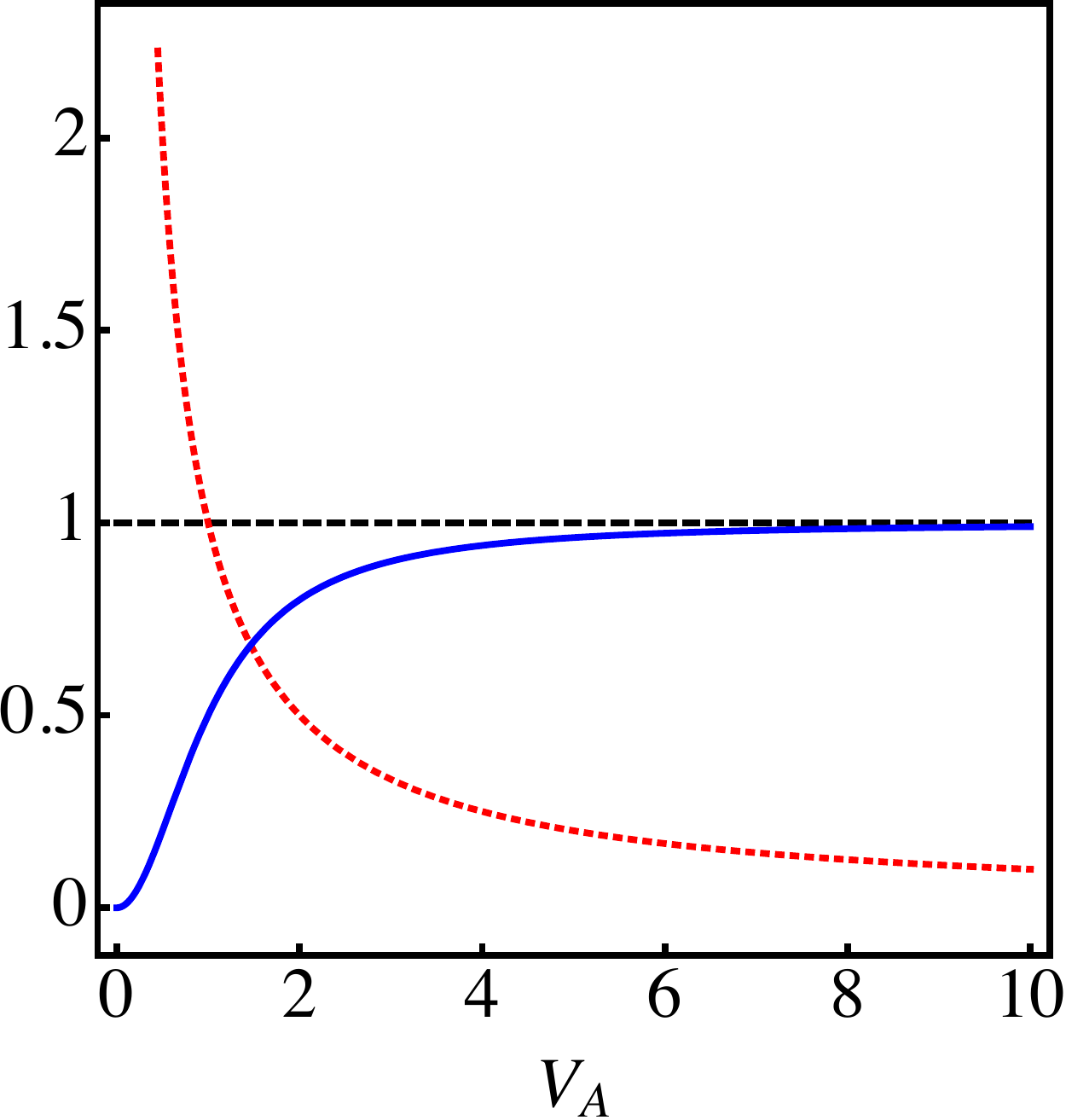}
\caption{In dotted red the ratio $|r_A|/|r|$ and in blue $|r|^2+|r_A|^2$ as a function of $V_A$ (in units of $\alpha$).}
\label{fig:reflection_coeff}
\end{figure}

In Fig.~\ref{fig:reflection_coeff} we plot $|r|^2+|r_A|^2$ and $|r_A|/|r|$ as a function of $V_A$, while setting $\alpha=1$. This shows that for large enough Andreev potentials the impurity models the NS junction asymptotically well, i.e., $|r|^2+|r_A|^2 \approx 1$. On the other hand, we see that with increasing $V_A$ the ratio between the Andreev and the regular reflection is decreasing, e.g., for $V_A=3.5$ for which  $|r|^2+|r_A|^2 \approx 0.92$, we barely have $|r_A|/|r| \approx0.3$. This corresponds to a value of $Z\approx 1$ for the NS junction (see Eq.~(\ref{eq:ZparameterNSjunction})). While the corresponding junction is definitely far from perfect Andreev reflection it can still support Andreev bound states, and we describe their behavior in Sec.~IV. This situation describes probably quite accurately the realistic parameters for many experimental  NS interfaces.

\section{\boldmath{$T$}-matrix formalism}
In order to take into account the cumulated effect of all-order impurity scattering processes, we employ the $T$-matrix formalism. We start with the case of a single impurity. Denoting the momentum-space Hamiltonian of a given system $\mathcal{H}_k$, we define the unperturbed Matsubara Green's function (GF) as: $G_0\left(k,i\omega_n\right) = \left[i\omega_n - \mathcal{H}_{k}\right]^{-1}$, where $\omega_n$ denote the Matsubara frequencies. In the presence of an impurity, the Green's function is modified to:
\begin{align}
\label{eq:green_perturbed}
G\left( k_1 ,k_2,i\omega_n \right) &= G_0\left(k_1,i\omega_n\right) \delta_{k_1,k_2} \\
&+G_0\left(k_1,i\omega_n\right) T\left(k_1,k_2,i\omega_n\right) G_0\left(k_2,i\omega_n\right), \nonumber 
\end{align}
where the $T$-matrix $T\left(k_1,k_2,i\omega_n\right)$ embodies impurity scattering processes \cite{Balatsky2006, Mahan2000}. For a delta-function impurity $V_{\mathrm{imp}}\left(x\right) = V \delta \left(x\right)$, the form of the $T$-matrix in 1D is momentum independent and is given by \cite{Balatsky2006, Ziegler1996, Salkola1996, Bena2008}:
\begin{equation}
\label{eq:T_matrix1D}
T\left(i\omega_n\right) = \left[\mathbb{I} - V \cdot \int \frac{dk}{2 \pi} G_0\left(k,i\omega_n\right)\right]^{-1} \cdot V.
\end{equation}
To compute physical quantities such as, e.g., the local density of states, in what follows we use the retarded GF $\mathcal{G}(k_1, k_2, E)$ obtained by the analytical continuation of the Matsubara GF $G(k_1, k_2, i\omega_n)$ (i.e., by setting $i\omega_n \rightarrow E + i\delta$, with $\delta \rightarrow 0^+$). The real space equivalents of Eqs.~(\ref{eq:green_perturbed}) and ~(\ref{eq:T_matrix1D}) can be written as
\begin{align*}
\mathcal{G}(x,x',E) = \mathcal{G}_0(x- x',E)+\mathcal{G}_0(x,E)T(E)\mathcal{G}_0(-x',E)
\end{align*}
and
\begin{align*}
T(E) &= \left[\mathbb{I} - V \cdot \mathcal{G}_0(x=0,E) \right]^{-1} \cdot V,
\end{align*}
respectively, where $\mathcal{G}_0(x=0,E)$ is equivalent to integrating the Green's function over all momenta (cf. Eq.~(\ref{eq:T_matrix1D})).

In the presence of two delta-function impurities with amplitudes $V_i$, localized at $x = X_i$, $i \in \{1,2\}$, i.e.,
\begin{align}
V_{\mathrm{imp}}\left(x\right) = V_1 \delta \left(x-X_1\right) + V_2 \delta \left(x-X_2\right),
\label{eq:twoAndreevImpPotentials}
\end{align}
the full $T$-matrix can be found from the following equation using the unperturbed retarded Green's function real-space form \cite{Choi2004,Choi2005}:
\begin{align}
T_{ij} = V_i \delta_{ij} \negthickspace + \negthickspace V_i \mathcal{G}_0(X_i\negthickspace-\negthickspace X_1)T_{1j} \negthickspace +\negthickspace V_i \mathcal{G}_0(X_i\negthickspace-\negthickspace X_2)T_{2j},
\label{eq:T_matrix_2imp}
\end{align}
where $i,j \in \{1,2\}$. For brevity we omit energy-dependence in all functions. Solving the system of four equations in Eq.~(\ref{eq:T_matrix_2imp}), we get:
\begin{align}
\nonumber T_{11} &= \left[\mathbb{I}\negthickspace -\negthickspace V_1 \mathcal{G}_0(0) \negthickspace-\negthickspace V_1 \mathcal{G}_0(X_1-X_2) T^{(0)}_2 \mathcal{G}_0(X_2-X_1)  \right]^{-1}\negthickspace V_1, \\
\nonumber T_{12} &= T^{(0)}_1 \mathcal{G}_0(X_1-X_2) T_{22}, \\
\nonumber T_{21} &= T^{(0)}_2 \mathcal{G}_0(X_2-X_1) T_{11}, \\
\nonumber T_{22} &= \left[\mathbb{I}\negthickspace -\negthickspace V_2 \mathcal{G}_0(0)\negthickspace -\negthickspace V_2 \mathcal{G}_0(X_2-X_1)T^{(0)}_1 \mathcal{G}_0(X_1-X_2)  \right]^{-1}\negthickspace V_2,
\end{align}
where we defined $T^{(0)}_i \equiv \left[\mathbb{I} - V_i \mathcal{G}_0(0)\right]^{-1} V_i$. The poles of the $T$-matrix $||T_{ij}||$ yield the energies of the bound states in the system. Furthermore, we can also express the full perturbed Green's function in terms of the $T$-matrix elements:
\begin{align}
\mathcal{G}(x,x') = \mathcal{G}_0(x\negthickspace-\negthickspace x') \negthickspace+\negthickspace \sum\limits_{ij}\mathcal{G}_0(x\negthickspace-\negthickspace X_i)T_{ij}\mathcal{G}_0(X_j\negthickspace-\negthickspace x').
\label{eq:perturbed_GF_2imp}
\end{align}
The correction to the local density of states due to the impurities can be expressed as 
\begin{align}
\Delta\rho(x) = -\frac{1}{\pi}\im \tr \sum\limits_{ij}\mathcal{G}_0(x\negthickspace-\negthickspace X_i)T_{ij}\mathcal{G}_0(X_j\negthickspace-\negthickspace x')\Bigg|_{x'=x}.
\label{eq:LDOS}
\end{align}

\section{Results}
The normal part of the junction is described using a simple lattice Hamiltonian
$
\mathcal{H}_{k} = \xi_k \tau_z,
$
where $\tau_z$ is the Pauli matrix acting in the particle-hole subspace, $\xi_k \equiv \mu - 2t \cos k a$, where $a$ is the lattice constant, $t$ denotes the hopping parameter and $\mu$ is the chemical potential. The retarded Green's function in real space is computed as the Fourier transform of its momentum-space representation, and is given by
\begin{align}
\mathcal{G}_0(E,x) = \bpm \mathcal{G}_{11}(E,x) & 0 \\ 0 & \mathcal{G}_{22}(E,x)\epm,
\end{align}
where 
\begin{align}
\label{eq:G11}&\mathcal{G}_{11}(E,x) = \\
\nonumber &- \frac{\left[\left(E_t-\mu_t +i\delta\right) - \sgn(E_t-\mu_t)\sqrt{\left(E_t-\mu_t +i\delta \right)^2 \negthickspace-\negthickspace 4}\right]^{\frac{|x|}{a}}}{at \cdot \sgn(E_t-\mu_t) \cdot 2^{\frac{|x|}{a}} \cdot\sqrt{\left(E_t-\mu_t +i\delta \right)^2 -4}}\\
\label{eq:G22}&\mathcal{G}_{22}(E,x) = \\
\nonumber &+ \frac{\left[\left( E_t+\mu_t +i\delta\right) - \sgn(E_t+\mu_t)\sqrt{\left(E_t+\mu_t +i\delta \right)^2-4}\right]^{\frac{|x|}{a}}}{at \cdot \sgn(E_t+\mu_t)\cdot 2^{\frac{|x|}{a}} \cdot \sqrt{\left(E_t+\mu_t +i\delta \right)^2-4}}
\end{align}
if $E^2+\mu^2 \neq 0$, and 
\begin{align}
\label{eq:G11x0} \mathcal{G}_{11}(E,x) &= -\frac{i}{2 a t} e^{+i\frac{\pi|x|}{2a}},\\
\label{eq:G22x0} \mathcal{G}_{22}(E,x) &= -\frac{i}{2 a t} e^{-i\frac{\pi|x|}{2a}},
\end{align}
for $E_t^2+\mu_t^2 = 0$. The expressions in Eqs.~(\ref{eq:G11x0}) and (\ref{eq:G22x0}) are obtained directly from the integral defining the Fourier transform of the momentum-space Green's function. In other words, we first set in the integrand $E_t = \mu_t = 0$ and then we perform the integral over momenta to obtain the real-space formulae. We expressed the energy and the chemical potential in terms of the hopping amplitude, i.e., $E_t \equiv E / t$ and $\mu_t = \mu / t$. The positive infinitesimal shift of energy, $+i\delta$, $\delta \to +0$, corresponds to an inverse quasiparticle lifetime and is generally related to the width of the energy levels. Note that since the expressions above are obtained within the lattice model, the results are valid only for $x = n a$, where $n \in \mathbb{Z}$.

The results that we present in this section are evaluated using this full tight-binding model. However, in order to test the validity of our approximations, we establish also a correspondence between the continuum model (used in Sec.~II to make the connection between an NS junction and an ``Andreev impurity'') and the lattice model. We thus expand $\xi_k$ in a quadratic form:
\begin{align}
\nonumber \xi_k &= \mu -2t \cos k a  \approx t a^2 k^2 + \mu - 2t = \frac{\hbar^2 k^2}{2m} - \frac{\hbar^2 k_F^2}{2m} .
\end{align}
This allows us to extract:
\begin{align}
\nonumber t &= \frac{\hbar^2}{2ma^2},\; \alpha \equiv \frac{\hbar^2 k_F^2}{m} = 2(2t-\mu ).
\end{align}
In what follows we use values of the chemical potential around $\mu=1.5$ so that $\alpha\approx 1$. We also set by default the hopping parameter $t=1$, therefore, making $\mu_t$ and $E_t$ equivalent to $\mu$ and $E$, respectively. The broadening is hereinafter set to $\delta = 0.001$.

To model the SNS junction we introduce two Andreev impurities, as in Eq.~(\ref{eq:AndreevImpPotential}). In order to take into account the phase difference between the superconductors we replace $V_A \to V_A e^{ \pm i\varphi}$ for one of the impurities, choosing the signs differently for the $12$ and $21$ components, to preserve the Hermitian character of the Hamiltonian. Using the notations from Eq.~(\ref{eq:twoAndreevImpPotentials}), we can thus write:
\begin{align}
V_1 = 
	\bpm
	0 & V_A \\
	V_A & 0
	\epm
	\;
	\text{and}
	\;
V_2 = 
	\bpm
	0 & V_A e^{-i\varphi} \\
	V_A e^{+i\varphi} & 0
	\epm.
\end{align}

Note that since we assume that the energy is much smaller than the SC gap our results apply to the long SNS junction limit. As mentioned in Sec.~II we focus on some intermediate value of $V_A \approx 3.5$ corresponding to a ratio of Andreev and regular reflection of $0.3$, and we study the formation of Andreev bound states using the formalism described in Sec.~III, the $T$-matrix for two delta-function impurities. This yields an exact formula for the perturbed Green's function of the system (see Eq.~(\ref{eq:perturbed_GF_2imp})), which allows to extract the local density of states (see Eq.~(\ref{eq:LDOS}))\footnote{Extra insight and a more detailed analysis of this exact result in the limit of the continuum model is provided in the Appendix}. We first plot the latter in Fig.~\ref{fig:LDOS} as a function of energy and position, for a fixed value of the chemical potential. Note the formation of bound states in the region between the two impurities, i.e., for $x \in \left[0,\,200 \right]$. 
\begin{figure}[t!]
\hspace{-0.5cm}
\includegraphics[width=6cm]{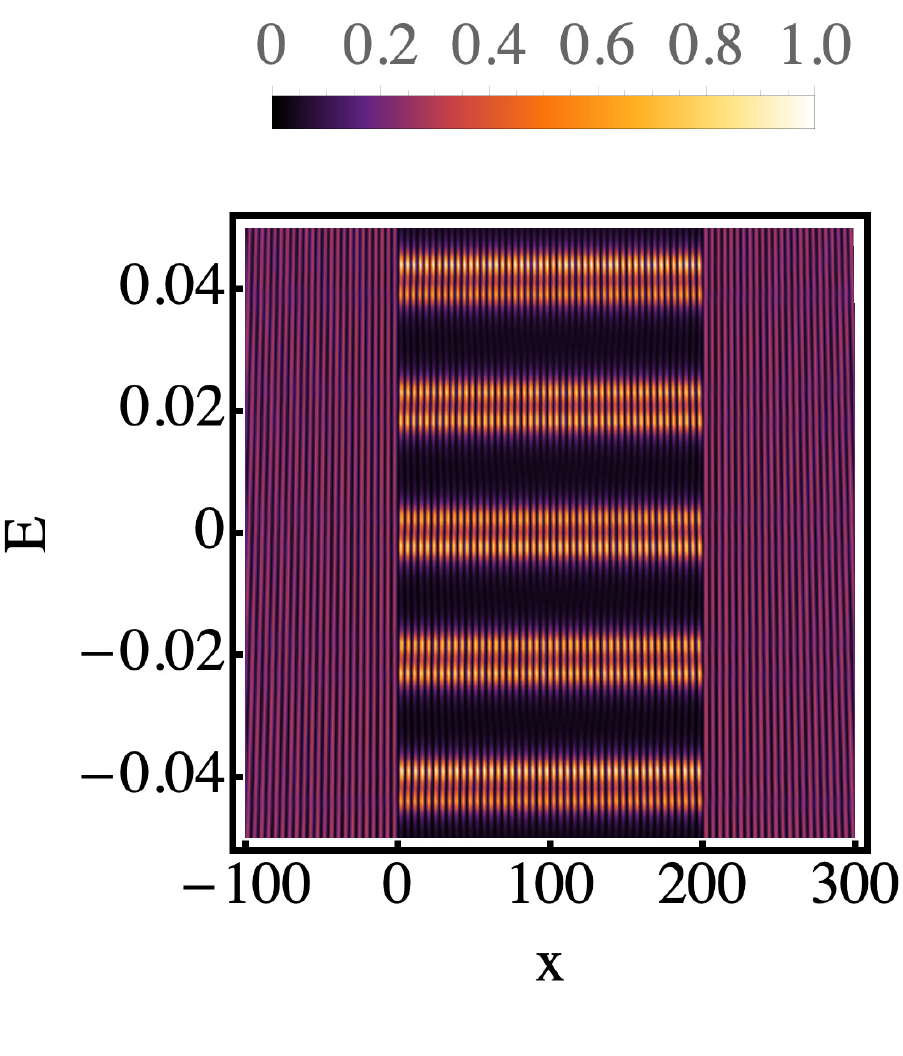}
\vspace{-0.5cm}
\caption{The LDOS of an SNS junction as a function of the coordinate $x$ and the energy $E$ on the horizontal and vertical axes, respectively. Two Andreev impurities are localized at $x=0$ and $x=200$. ABS form in the normal region. The length of the normal region is $L=200$, the chemical potential is taken to be $\mu=1.5$, $V_A=3.5$ and the phase difference $\varphi=0$.}
\label{fig:LDOS}
\end{figure}

\begin{figure}[ht!]
\hspace{-0.5cm}
\includegraphics[width=10cm]{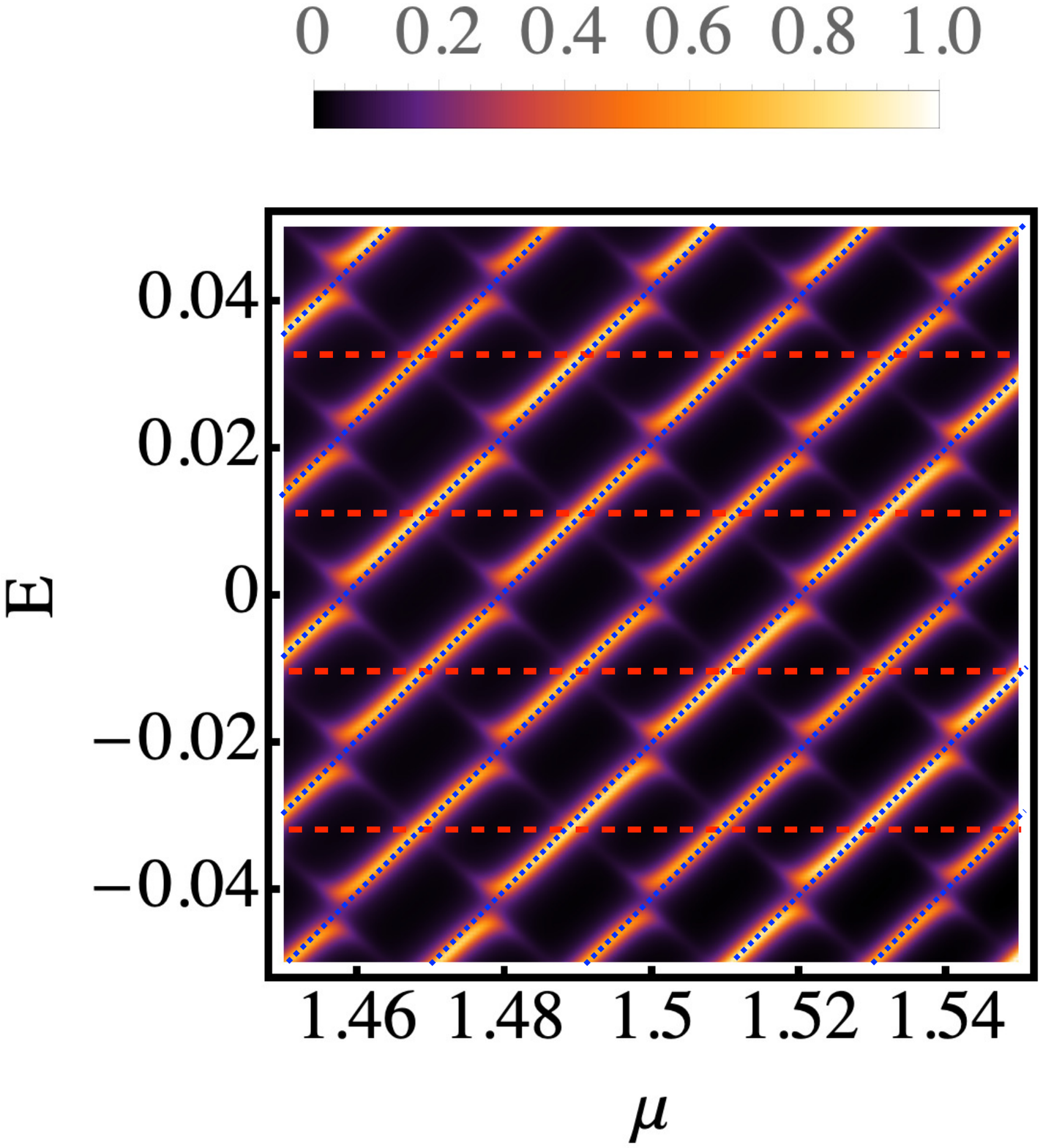}
\vspace{-0.5cm}
\caption{LDOS averaged over 6 sites $x \in \left[90,\,95\right]$, plotted as a function of the chemical potential on the horizontal axis and energy on the vertical axis. We set $L=200$, $V_A=3.5$ and $\mu$ varies between $1.46$ and $1.54$, so that $\alpha\approx 1$ does not vary with $\mu$. The phase difference is set to $\varphi=0$. The perfect Andreev limit is denoted by the red dashed lines while the particle-in-the-box by the blue dotted lines.}
\label{fig:chemical_potential}
\end{figure}

In order to compare the behavior of these bound states with that of previously studied ABS \cite{Bena2012,Pillet2010,Dirks2011}, we focus on a given position and we plot in Fig.~\ref{fig:chemical_potential} the dependence of the LDOS as of function of energy and chemical potential. In the plots we choose to average the LDOS over a few sites (i.e., 6 sites) to avoid any fluctuation effects due to the short wavelength oscillations visible in Fig.~\ref{fig:LDOS}. Indeed, we recover the oscillatory bell-shaped dependence of the ABS energy on the chemical potential (see, e.g., Figs.~4a and 4d in Ref.~[\onlinecite{Bena2012}]).

\begin{figure}[ht!]
\hspace{-0.5cm}
\includegraphics[width=10cm]{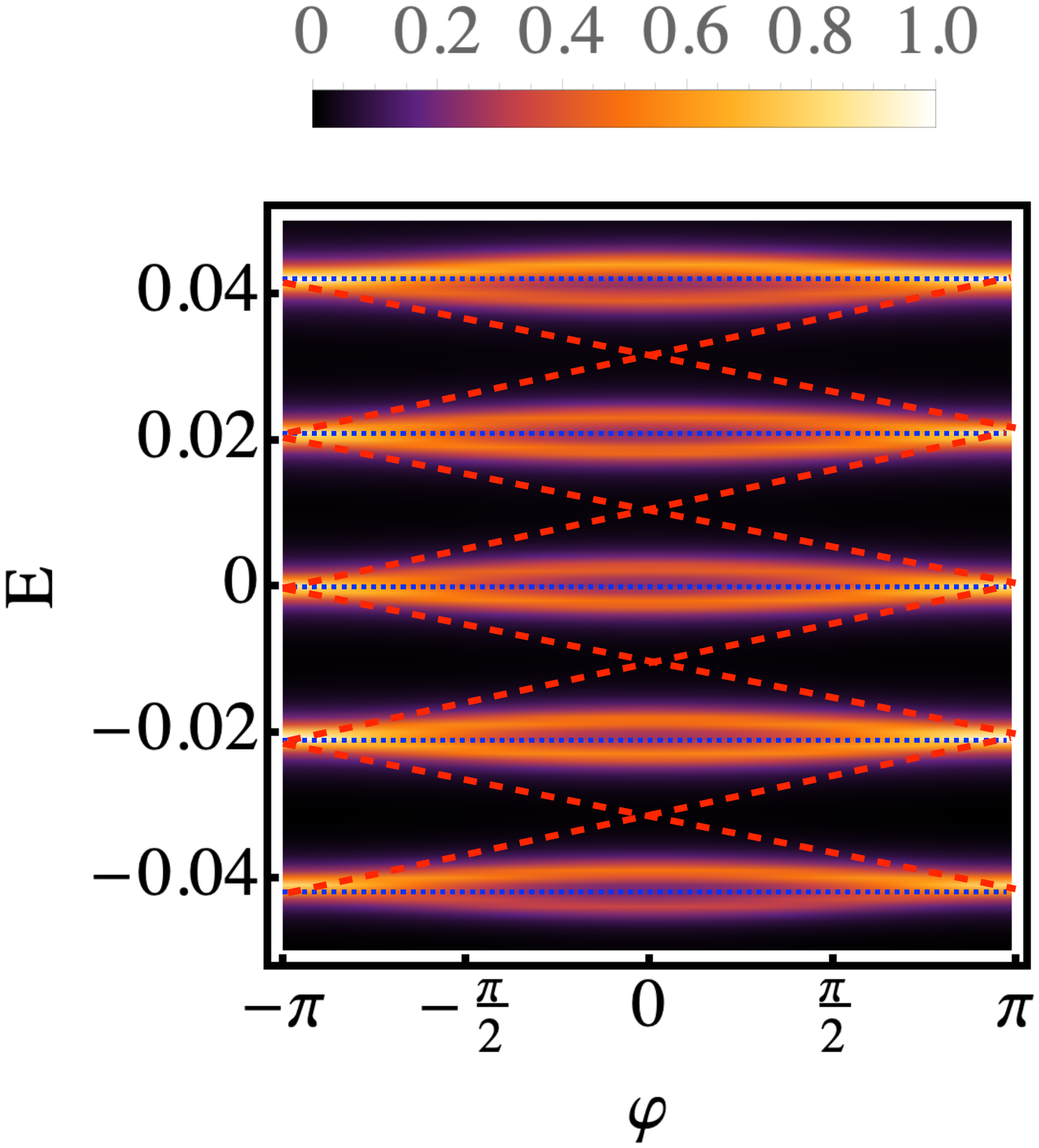}
\vspace{-0.5cm}
\caption{LDOS averaged over 6 sites $x \in \left[90,\,95\right]$, plotted as a function of the phase difference $\varphi$ on the horizontal axis and energy on the vertical axis. We set $L=200$, $V_A=3.5$ and $\mu=1.5$. The perfect Andreev limit is denoted by the red dashed lines while the particle-in-the-box by the blue dotted lines.}
\label{fig:phase}
\end{figure}

Furthermore, in Fig.~\ref{fig:phase} we plot the dependence of the LDOS as a function of energy and the phase difference between the two SCs, for a fixed chemical potential. We note that the amplitude of the phase oscillations, while not very large, is still significant, marking the presence of nonzero Andreev reflection and of the corresponding phase coherence. 

In order to understand this observed dependence of the ABS energies on the phase and chemical potential we remind the reader that the focus of our work is the study of long SNS junctions in the \textit{imperfect} long junction limit. Also we focus on the limit of small energies with respect to the gap, $E \ll \Delta$. In what follows we write down the solutions for the two extreme limits characterizing this junction, that have both been studied previously in the literature. First, for the long perfect SNS junctions (no normal backscattering at the leads), the corresponding results are described by the Andreev approximation. In this limit the energies of the ABS are linear with the phase and independent of chemical potential, and they are given by
\begin{equation}
E_n^{\pm A}=\pm \frac{\hbar p_F}{2 m L} [\phi-(2n+1)\pi], \; n \in \mathbb{N},
\end{equation}
To the other extreme, in the limit of zero Andreev scattering and infinite normal scattering at the leads our setup is equivalent to the particle-in-a-box problem, which for quadratically-dispersed quasiparticles has the following well-known solution:
\begin{equation}
E_n^{\pm B} = \pm \left[\frac{\hbar^2  \pi^2 n^2}{2 m L^2} - \frac{p_F^2}{2 m} \right], \; n \in \mathbb{N},
\end{equation}
independent of phase, and linear with the chemical potential.

To make the correspondence with the lattice-model results we choose $m=0.5$, and we express $p_F$ in terms of the chemical potential $\mu$, using the previously introduced definition of $\alpha \equiv p_F^2/m = 2(2t-\mu)$. Since we have taken $t=1$ we need to set $p_F = \sqrt{2-\mu}$. 
Thus we can rewrite
\begin{align}
E_n^{\pm A} &= \pm \frac{1}{\sqrt{2} L}\left[\phi-(2n+1)\pi\right], \; n \in \mathbb{N}, \\
\label{eq:boxenergiesmu} E_n^{\pm B} &= \pm \left[\frac{\pi^2 t  n^2}{L^2} - 2t + \mu \right], \; n \in \mathbb{N},
\end{align}
where $t=1$ and $L$ is expressed in units of the lattice constant $a$.

The corresponding limits for the bound states in the junction are depicted in Figs.~\ref{fig:chemical_potential} and \ref{fig:phase} by the dashed red lines (Andreev limit) and the dotted blue lines (particle-in-a-box limit). Note that our results interpolate between the two regimes; our approach is unique in the sense that it is the only analytical approach to capture this transition between a long SNS junction with perfect contacts and a perfectly isolated system. Moreover, as we will show in what follows, by expanding about the particle-in-the-box limit, we can obtain a fully closed form for the ABS dependence of energy depicted in Figs.~\ref{fig:chemical_potential} and \ref{fig:phase}, valid for all the values of the chemical potential with the exception of a countable set of points.

\begin{figure}[ht!]
	\centering
	\includegraphics[width=9cm]{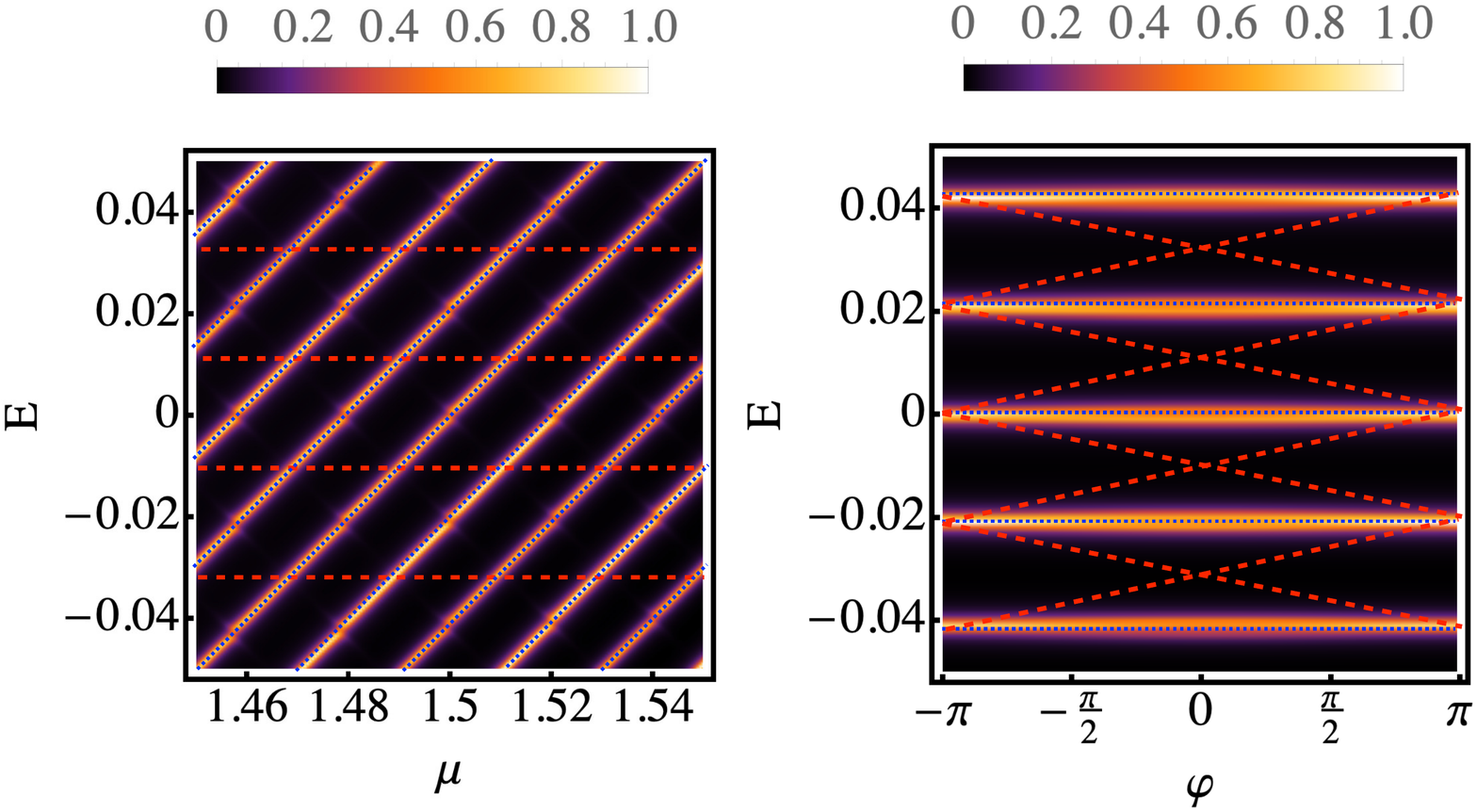}
	\vspace{-1cm}
	\caption{LDOS averaged over 6 sites $x \in \left[90,\,95\right]$, plotted as a function of energy (vertical axis) and on the horizontal axis the chemical potential (left panel) and the SC phase difference (right panel). Here we set $L=200$, $V_A = 10$. In the left panel we fix  $\varphi=0$, whereas in the right panel $\mu=1.5$. The perfect Andreev limit is denoted by the dashed red lines while the particle-in-the-box by the dotted blue lines.}
	\label{fig:LDOSimperfect1}
\end{figure}

We also plot the dependence of the ABS on chemical potential and phase difference for parameter values that take us closer to the two ideal limits, for example we take a smaller $|r_A|/|r|$ (for example, take $V_A = 10$) such that we recover a junction  very close to the particle-in-the-box scenario (see Fig.~\ref{fig:LDOSimperfect1}). 

Note that in this limit the Andreev reflection goes to zero and the bell-shaped oscillations transform into almost fully-linear crossings (reflecting the linear energy dependence of the bound states in a quantum dot on gate voltage, corresponding to Eq.~(\ref{eq:boxenergiesmu}) and denoted in blue). Moreover, the dependence of the ABS energy on the SC phase difference becomes almost insignificant, as expected.

\begin{figure}[ht!]
	\centering
	\hspace{-0.5cm}
	\includegraphics[width=9cm]{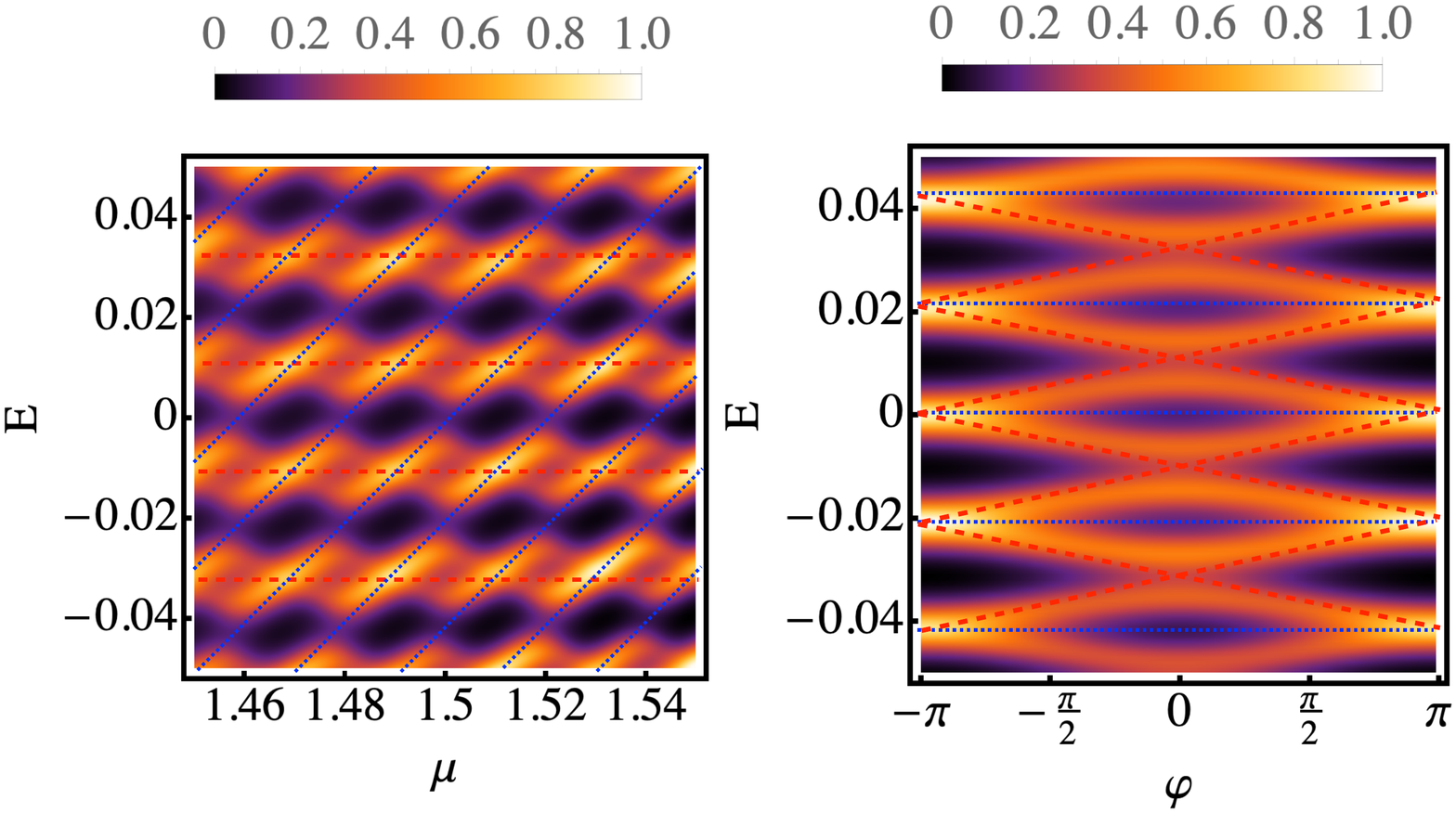}
	\vspace{-1cm}
	\caption{LDOS averaged over 6 sites $x \in \left[90,\,95\right]$, plotted as a function of energy (vertical axis) and on the horizontal axis the chemical potential (left panel) and the SC phase difference (right panel). Here we set $V_A = 1$. The other parameters are the same as in Fig.~\ref{fig:LDOSimperfect1}. The perfect Andreev limit is denoted by the red dashed lines while the particle-in-the-box by the blue dotted lines.}
	\label{fig:LDOSimperfect2}
\end{figure}

On the other hand, we focus on smaller $V_A$ values (for instance, $V_A=1$), artificially increasing $|r_A|/|r|$, however  the results we obtain are no longer physical and consistent with the NS junction since $|r_A|^2+|r|^2 \ll 1$ (see Fig.~\ref{fig:LDOSimperfect2}). Nevertheless, the main features of the ABS are generally preserved, and we see that we get closer to the Andreev limit denoted in red.

Moreover, we can obtain a fully closed form for the ABS energy levels if we employ the continuum model introduced in Sec.~II, namely we consider a system with quadratic dispersion (see the Appendix) and expand around the particle-in-the-box-limit. The zeroth order contribution to the ABS level is thus given by Eq.~(\ref{eq:boxenergiesmu}). Subsequently we calculate the first-order correction to this solution in terms of the small dimensionless parameter $(p_F/m V_A)^2$: 
\begin{align}
\label{eq:correction}\delta E_n^{\pm B} &= \mp \frac{2\pi^2 n^2}{i p_F^3 L^3}  \left(\frac{p_F}{m V_A}\right)^2 \sqrt{2 - \frac{\pi^2n^2}{p_F^2 L^2}}\\
\nonumber & \times\frac{1+(-1)^n \cos\varphi\, \exp\left(ip_F L \sqrt{2 - \frac{\pi^2n^2}{p_F^2 L^2}}\right) }{1-\exp\left(2ip_F L \sqrt{2 - \frac{\pi^2n^2}{p_F^2 L^2}}\right)} \frac{p_F^2}{2m}.
\end{align}
We are leaving the technical details of the derivation to the Appendix. This correction to the particle-in-a-box energies has both real and imaginary parts. The real part is responsible for the energy shifts, whereas the imaginary part embodies the energy level broadening. 

\begin{figure}
	\centering
	\includegraphics[width=1.2\columnwidth]{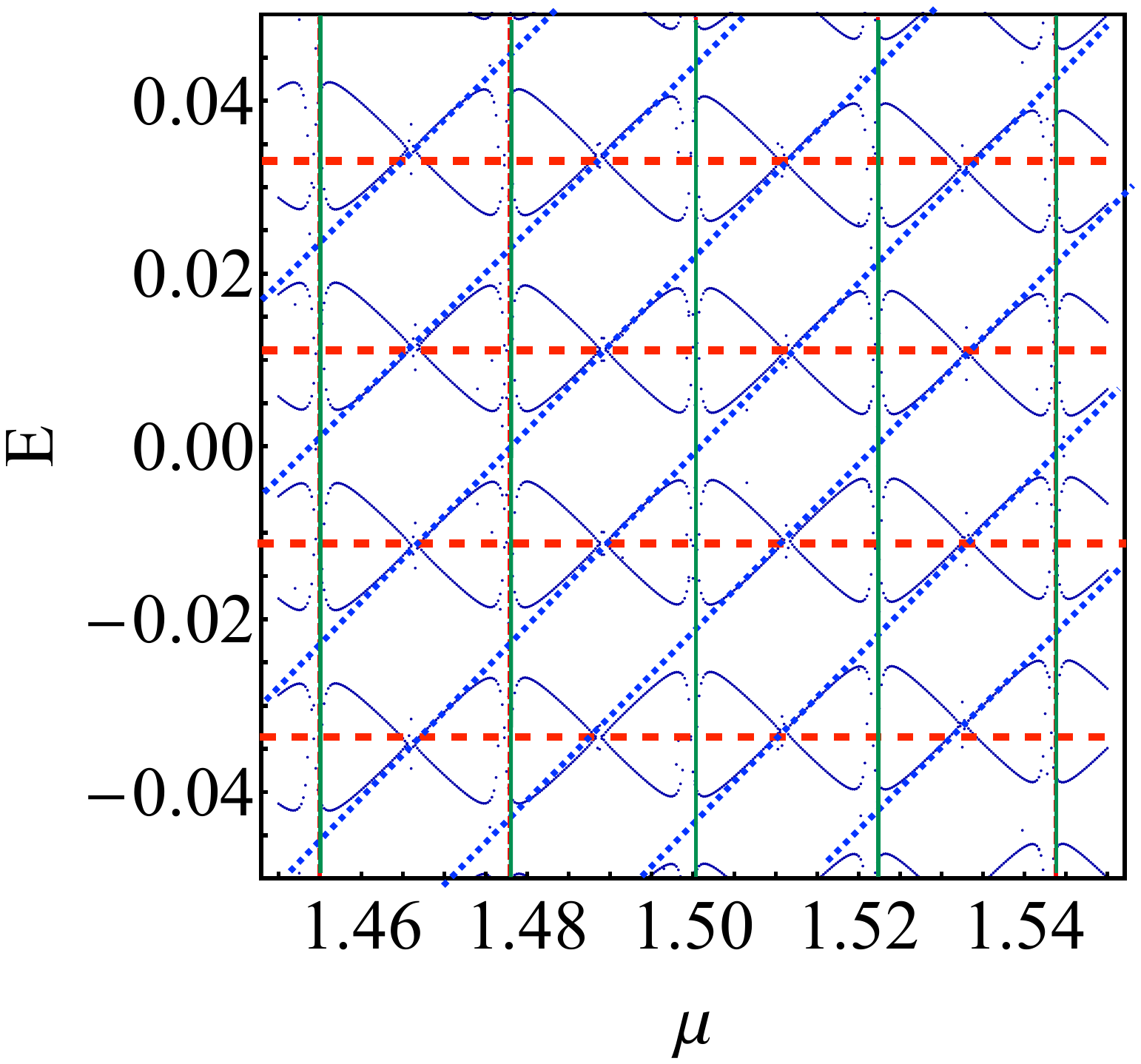}
	\caption{Energies of the ABS within the continuum model plotted as a function of the chemical potential. The vertical green lines mark the values of the chemical potential at which the analytical perturbative approach becomes invalid due to singularities in Eq.~(\ref{eq:correction}). Here, exactly as in Fig.~\ref{fig:chemical_potential}, we have set $V_A = 3.5$, $\varphi=0$ and $L=200$. Additionally, we have fixed $m=0.5$. The perfect Andreev limit is denoted by the red dashed lines while the particle-in-the-box by the blue dotted lines.}
	\label{fig:ABSenergiesanalyt}
\end{figure}

Eq.~(\ref{eq:correction}) where we set $p_F = \sqrt{2-\mu}$ allows us to plot the energies $E_n^{\pm B} + \delta E_n^{\pm B}$ as a function of the chemical potential $\mu$ (see Fig.~\ref{fig:ABSenergiesanalyt}). There is a very good agreement between these closed-form analytical results for the continuum model and those presented in Fig.~\ref{fig:chemical_potential} corresponding to a direct evaluation of the T-matrix and the corresponding LDOS for a full tight-binding model. Note that our perturbative treatment within the continuum model does not work for values of the chemical potential close to the singularities in Eq.~(\ref{eq:correction}), which constitute a countable set of points. 

A subsequent simplification of Eq.~(\ref{eq:correction}) allows us to extract some other important information about the ABS level behavior. First, we recall that our approach is valid at small energies, $\delta E^{\pm B}_n/\mu \ll 1$. Hence we can take
$\frac{\pi^2 n^2}{p_F^2 L^2} \approx 1$, and $\sqrt{2 - \frac{\pi^2n^2}{p_F^2 L^2}} \approx 1$ in  Eq.~(\ref{eq:correction}). Thus, we can extract the periodicity of the Andreev levels to be $\delta p_F \cdot L\approx 2\pi$ and consequently $|\delta \mu| = \frac{4\pi}{L} \sqrt{2-\mu}$. For the values considered here ($L=200$, $\mu\approx 1.5$) we see that this corresponds to $\delta \mu=0.045$, which is observed in both Fig.~\ref{fig:chemical_potential} and Fig.~\ref{fig:ABSenergiesanalyt}. On the other hand, the singularities in Eq.~(\ref{eq:correction}) are governed by  $2 \delta p_F L\approx 2\pi$, yielding $|\delta \mu|=0.022$ corresponding to the spacing between the green vertical lines in Fig.~\ref{fig:ABSenergiesanalyt}.

Moreover, Eq.~(\ref{eq:correction}) allows us to see that the amplitude of the phase oscillations, i.e., the coefficient of the $\cos(\varphi)$ term, is proportional to $(p_F/m V_A)^2$. Indeed, this goes to zero when there is no Andreev reflection $V_A\rightarrow \infty$, and increases with reducing $V_A$. Eqs.~(\ref{eq:ZparameterNSjunction}) and (\ref{eq:rA}) allow to make a direct correspondence between the amplitude of these oscillations and the dimensionless barrier strength $Z$, namely, it is decreasing roughly as $1/Z^4$.

Furthermore, the width of the ABS levels can also be extracted from the imaginary part of Eq.~(\ref{eq:correction}), and is thus equal to $\frac{1}{p_F L}  \left(\frac{p_F}{m V_A}\right)^2\approx 0.001$ for $V_A=3.5$, consistent with Figs.~\ref{fig:chemical_potential} and \ref{fig:phase}. Note that with decreasing $V_A$ these estimates become inaccurate and the higher order terms become important (see Appendix A), for example, for $V_A=3.5$ the next order correction which we neglect here is of the order of $\approx 16\%$.

Note that the broadening of the ABS levels is governed by two physical phenomena. The first is the finite quasiparticle lifetime, which is quite large in regular systems making the levels very sharp. We introduce by hand an artificial broadening, described by the imaginary contribution $i \delta$, with $\delta=0.001$, into the energy in the Green's functions. It is proportional to the inverse quasiparticle lifetime and it is the main factor of level broadening in regular systems. However, in a setup consisting of a central region coupled to the leads, there is a second physical factor responsible for the level broadening, originating from the coupling to the leads. When the coupling to the leads is purely Andreev-like, the width of the ABS is not affected by its size, see Fig.~7 in Ref.~[\onlinecite{Bena2012}]. However, when there is also a normal coupling we expect that the ABS level width is going to be affected by this normal coupling, same as the quantum dot levels becoming wider when the dot is better coupled to the leads; in the perfect coupling limit such levels can become very wide, while in the bad coupling limit the levels are very sharp (see for example Figs.~4a and 4b in Ref.~[\onlinecite{Bena2012}]  for details). In our model, it is the parameter $V_A$ that determines the barrier height and thus the coupling to the SC leads. Therefore, we expect that if the coupling is very small, i.e., $V_A \to \infty$, then the energy level broadening is controlled solely by $\delta$ and we obtain the `particle in a box solutions'. Indeed, this is exactly what happens (see Eq.~(A13-A15) in Appendix A), and in this limit the levels are very sharp. On the other hand, when $V_A$ is small we have an additional broadening due to the normal coupling to the leads and this is indeed what we observe.

We should note that no alternative simple expression for our Eq.~(\ref{eq:correction}) providing a closed form for the energies of the ABS levels as a function of phase difference and chemical potential, has ever been derived for long SNS junctions with imperfect contacts. This demonstrates the strength of our approach to obtain analytical insight into problems for which there is no other analytical alternative.

\section{Conclusions}

We have shown that by introducing two ``Andreev'' impurities into a normal metal and by employing the $T$-matrix formalism, we can model a long imperfect SNS junction and the formation of Andreev bound states. Most importantly, we have obtained analytical expressions of the local density of states, as well as of the energies of the ABS on different parameters of the system, including the chemical potential and the phase difference between the SCs.  In our model the Andreev to normal scattering ratio can be controlled by varying the Andreev impurity potential amplitude. We show that our method interpolates between the Andreev limit corresponding to a perfect long SNS junction for which the energies of the ABS are independent of chemical potential and linear with phase, and the particle-in-the-box limit in which the energy levels are independent of phase and linear with chemical potential. One of the most spectacular results recovered by this technique is a closed-form expression of the energies of the ABS levels, as well as of their widths, obtained by a first-order expansion around the particle-in-the-box limit, as a function of the chemical potential, the SC phase difference, and the interface barrier strength, till now only possible by non-analytical tools such as numerical tight-binding exact diagonalization.
This makes our method a unique analytical tool to understand and characterize ABS. Moreover, our technique can be easily generalized to other more complex setups, for example, systems with asymmetric leads, or in which the spin of the electrons also plays a role. We have checked for example that in our formalism the Zeeman field only splits the energies of the ABS levels, consistent with an intuitive understanding. However, more complex factors such as Rashba spin-orbit coupling would require a modification of the equations of motion, and of all the conditions for the equivalence with the Andreev impurity; if successful, such an analysis will be addressed in a separate work.

Our approach thus provides an easily accessible non-numerical tool to describe the formation of ABS in long SNS junctions. While at present ABS have been directly observed only in short junctions---quantum-dot-type setups for which only a few states can be observed inside the SC gap---we can expect that longer and cleaner wires will also be realized in the future, for which models such as the one presented here will be necessary in order to investigate the underlying physics. Already some progress has been made in this direction\cite{Beenakker2013}, and some experimental setups in which a short but non-zero length plays a role have been examined\cite{Tosi2019}. In order to make contact with such experiments our approach may eventually need to be extended to include other factors, such as the presence of interactions and the spin-splitting of the levels.

\begin{acknowledgments}
V.K. would like to acknowledge the ERC Starting Grant No. 679722 and the Roland Gustafsson foundation for theoretical physics.
\end{acknowledgments}

\bibliography{biblio_ABS_Tmatrix}

\widetext
\appendix

\section{Andreev bound states in the continuum model}

We consider a 1D spinless normal metal with two Andreev impurities localised at $x=0$ and $x=L$. We can write down a simple model in the basis of electrons and holes as follows:
\begin{align}
\mathcal{H}= \mathcal{H}_{\mathrm{met}} + V_{\mathrm{imp}}(x) \equiv \xi_p \tau_z + V_1 \delta(x) + V_2 \delta(x-L), \quad\text{where}\quad \xi_p \equiv \frac{p^2}{2m}-\frac{p_F^2}{2m}.
\label{eq:AppHfull}
\end{align}
Above $p_F$ denotes the Fermi momentum, $m$ is the effective mass of electrons, $V_{1,2}$ are the amplitudes of the impurity potentials. The Andreev impurities in this model can be written as
\begin{align}
V_1 = \bpm 0 & V_A \\ V_A & 0\epm \quad\text{and}\quad V_2 = \bpm 0 & V_A e^{-i\varphi} \\ V_A e^{+i\varphi} & 0\epm,
\end{align} 
where $\varphi$ corresponds to the phase difference between the SCs. Such impurity potentials reflect electrons into holes, and vice versa. In what follows we study the resulting impurity-induced bound states. We start by computing the unperturbed retarded Green's function of the superconductor. In momentum space it is given by:
\begin{equation}
\mathcal{G}_0(E,p) \equiv \left[E-\mathcal{H}_0(p)\right]^{-1} = \frac{1}{ E^2 - \xi_p^2}
	\bpm 
		E+\xi_p & 0 \\
		0 & E-\xi_p 
	\epm
\end{equation}
To solve the problem we need to Fourier-transform the Green's function to real space as follows
\begin{equation*}
\mathcal{G}_0(E,x) = \int\frac{dp}{2\pi} \mathcal{G}_0(E,p)e^{ipx}
\end{equation*}
Two integrals are sufficient to define the real-space form of the Green's function:
\begin{align}
X_\pm(x) \equiv \int\frac{dp}{2\pi} \frac{e^{ipx}}{E \pm \xi_p +i0} = \pm i m \frac{e^{i\sqrt{p_F^2 \mp 2mE}\,|x|}}{\sqrt{p_F^2 \mp 2mE}}
\end{align} 
The infinitesimal shift in the denominator originates from the definition of the retarded Green's function. In terms of the functions defined above the Green's function in coordinate space becomes
\begin{align}
\mathcal{G}_0(E,x) =
	\bpm 
		- i m \frac{e^{i\sqrt{p_F^2 + 2mE}\,|x|}}{\sqrt{p_F^2 + 2mE}} & 0 \\
		0 & + i m \frac{e^{i\sqrt{p_F^2 - 2mE}\,|x|}}{\sqrt{p_F^2 - 2mE}}
	\epm
\end{align}

Now we can proceed to solving the Schr\"odinger equation for the Hamiltonian in Eq.~(\ref{eq:AppHfull}):
\begin{equation}
\left[ \mathcal{H}_{\mathrm{met}} + V_1 \delta(x) + V_2 \delta(x-L) \right] \Psi(x)= E \Psi(x)
\end{equation}
In the Fourier space we get:
\begin{align}
\Psi(p) =  \mathcal{G}_0(E,p) \cdot V_1  \cdot\Psi(x=0) + \mathcal{G}_0(E,p) e^{-ipL} \cdot V_2  \cdot \Psi(x=L)
\end{align}
Going back to the real space we have:
\begin{align}
\Psi(x) =  \mathcal{G}_0(E,x) \cdot V_1  \cdot\Psi(x=0) + \mathcal{G}_0(E,x-L)  \cdot V_2  \cdot \Psi(x=L)
\label{eq:AppWFrealspace}
\end{align}
First, we need to find the wave function values at $x=0$ and $x=L$, and then define the full coordinate dependence using those. Thus, we have a system of equations to solve (in what follows we omit $x=0,L$ in the argument of the wave function and just write $0,L$):
\begin{align}
\Psi(0) &= \mathcal{G}_0(E,0)  \cdot V_1  \cdot\Psi(0) + \mathcal{G}_0(E,-L)  \cdot V_2  \cdot \Psi(L) \\
\Psi(L) &= \mathcal{G}_0(E,L)  \cdot V_1  \cdot\Psi(0) + \mathcal{G}_0(E,0)  \cdot V_2  \cdot \Psi(L)
\end{align}
Or rewritten in a block-matrix form:
\begin{align}
\bpm  
\mathbb{I} - \mathcal{G}_0(E,0)  \cdot V_1 & - \mathcal{G}_0(E,-L)  \cdot V_2   \\
-\mathcal{G}_0(E,L)  \cdot V_1 & \mathbb{I} - \mathcal{G}_0(E,0)  \cdot V_2  
\epm 
\bpm \Psi(0) \\ \Psi(L) \epm = 0
\end{align}
In order to ensure that this equation has non-trivial solutions we set the determinant of the block matrix to zero, which in turn yields the equation for the energies of bound states:
\begin{align}
\nonumber\det \bpm  
\mathbb{I} - \mathcal{G}_0(E,0)  \cdot V_1 & - \mathcal{G}_0(E,-L)  \cdot V_2  \\
-\mathcal{G}_0(E,L)  \cdot V_1 & \mathbb{I} - \mathcal{G}_0(E,0)  \cdot V_2 
\epm = \phantom{aaaaaaaaaaaaaaaaaaaaaaaaaaaaaaaaaaaaaaaaaaaaaa}\\
= 1 - 2 \frac{m^2V_A^2}{p_F^2} \frac{1+ \cos \varphi \cdot e^{ip_FL \left(\sqrt{1-\gamma} + \sqrt{1+\gamma} \right)}}{\sqrt{1-\gamma^2}} + \frac{m^4V_A^4}{p_F^4} \frac{\left(1-e^{2ip_F L \sqrt{1-\gamma}} \right)\left(1-e^{2ip_F L \sqrt{1+\gamma}} \right)}{1-\gamma^2} = 0,
\label{eq:AppABSenergies}
\end{align}
where we have introduced a dimensionless parameter $\gamma = 2mE/p^2_F \equiv E/E_F$. The equation above is transcendental, and in the most general case does not have an analytical solution.\\

First, we turn to the limit of $V_A \to \infty$, in which Eq.~(\ref{eq:AppABSenergies}) is easily solvable: 
\begin{align}
\left(1-e^{2ip_F L \sqrt{1\pm \gamma^\pm_0}} \right) = 0 \;\Rightarrow\; p_F L \sqrt{1 \pm \gamma^\pm_0} = \pi n \;\Rightarrow\; E_n^\pm = \pm \left[ \frac{\pi^2n^2}{2m L^2} - E_F \right], \quad\text{where}\; n \in \mathbb{N}.
\label{eq:AppABSenergiesInfVA}
\end{align}
There is nothing surprising about this result: we have just obtained the energy levels of a particle in a box (since the limit of infinite potential corresponds to that). What is left is to find the wave functions at $x=0$ and $x=L$ in order to obtain the final expression for the wave function in the limit of $V_A \to \infty$. This can be done straightforwardly using Eqs.~(\ref{eq:AppWFrealspace}) while plugging in the energies obtained in Eq.~(\ref{eq:AppABSenergiesInfVA}):
For electrons we have:
\begin{align}
E_n^+ = + \left[ \frac{\pi^2n^2}{2m L^2} - E_F \right], \quad \Psi(0) = \bpm 0 \\ 1 \epm,\; \Psi(L) =\bpm 0 \\ (-1)^{n+1} \epm,\quad \Psi(x) = \bpm \frac{i m L}{\pi n} \left[e^{\frac{i\pi n |x|}{L}} + (-1)^{n+1} e^{\frac{i\pi n |x-L|}{L}} \right] \\ 0 \epm,
\end{align}
whereas for holes we obtain:
\begin{align}
E_n^- = - \left[ \frac{\pi^2n^2}{2m L^2} - E_F \right], \quad \Psi(0) = \bpm 1 \\ 0 \epm,\; \Psi(L) = \bpm (-1)^{n+1} \\ 0 \epm,\quad \Psi(x) =\bpm 0 \\ \frac{i m L}{\pi n} \left[e^{\frac{i\pi n |x|}{L}} + (-1)^{n+1} e^{\frac{i\pi n |x-L|}{L}} \right] \epm.
\end{align}
It is easy to verify that these wave functions correspond to those of a particle in a box.\\

%

In order to extract the Fermi energy dependence of Andreev bound state energies, we need to find the first-order correction to the particle-in-a-box solutions in Eq.~(\ref{eq:AppABSenergiesInfVA}). Hence, we rewrite Eq.~(\ref{eq:AppABSenergies}) in the following form:
\begin{align}
\left( 1-e^{2ip_F L \sqrt{1-\gamma}} \right)\left( 1-e^{2ip_F L \sqrt{1+\gamma}} \right) = \frac{1-\gamma^2}{\left(m V_A / p_F\right)^2}\left[2 \frac{1+\cos\varphi\, e^{ip_F L (\sqrt{1-\gamma}+\sqrt{1+\gamma})}}{\sqrt{1-\gamma^2}} - \frac{1}{\left(m V_A / p_F\right)^2} \right].
\label{eq:AppABSenergiesrewritten}
\end{align}
For the parameter range we are interested in this equation can be considered perturbatively with $p_F/m V_A$ being the small parameter. In the $0$-th approximation we have the same solutions as those for a particle in a box (see Eq.~(\ref{eq:AppABSenergiesInfVA})):
\begin{align}
\gamma^\pm_0 = \pm \left[\frac{\pi^2n^2}{p_F^2 L^2} - 1\right], \; n \in \mathbb{N}
\end{align}
In what follows we find the first correction to these two series of solutions. We demonstrate it using $\gamma = \gamma_0^+$. First, we divide Eq.~(\ref{eq:AppABSenergiesrewritten}) by $\left( 1-e^{2ip_F L \sqrt{1-\gamma}} \right)$, since by our choice of root this factor is never equal to zero: 
\begin{align}
 1-e^{2ip_F L \sqrt{1+\gamma}}  = \frac{1-\gamma^2}{\left(m V_A / p_F\right)^2} \frac{1}{1-e^{2ip_F L \sqrt{1-\gamma}}}\left[2 \frac{1+\cos\varphi\, e^{ip_F L (\sqrt{1-\gamma}+\sqrt{1+\gamma})}}{\sqrt{1-\gamma^2}} - \frac{1}{\left(m V_A / p_F\right)^2} \right]
\label{eq:Apptosolve}
\end{align}
We substitute the $0$-th approximation solution into the right-hand side of the Eq.~(\ref{eq:Apptosolve}). The left-hand side we represent as Taylor series around the point $\gamma = \gamma_0^+$ up to the first-order correction, i.e.,
\begin{align}
 1-e^{2ip_F L \sqrt{1+\gamma}} \approx  1-e^{2ip_F L \sqrt{1+\gamma_0^+}} + \frac{d(1-e^{2ip_F L \sqrt{1+\gamma}})}{d\gamma}\Big|_{\gamma = \gamma_0^+}\; \delta\gamma^+ = - \frac{i p_F^2 L^2}{\pi n} \delta\gamma^+,
\end{align}
where $\delta\gamma^+$ is the sought-for correction to $\gamma_0^+$. Computing the derivative and using Eq.~(\ref{eq:Apptosolve}) we get:
\begin{align}
\delta\gamma^+ = - \frac{\pi n}{i p_F^2 L^2} \frac{\frac{\pi^2n^2}{p_F^2 L^2} \left(2 - \frac{\pi^2n^2}{p_F^2 L^2} \right)}{\left(m V_A / p_F\right)^2} \frac{1}{1-\exp\left(2ip_F L \sqrt{2 - \frac{\pi^2n^2}{p_F^2 L^2}}\right)} \left[2 \frac{1+(-1)^n \cos\varphi\, \exp\left(ip_F L \sqrt{2 - \frac{\pi^2n^2}{p_F^2 L^2}}\right)}{\frac{\pi n}{p_F L}\sqrt{2 - \frac{\pi^2n^2}{p_F^2 L^2}}} - \frac{1}{\left(m V_A / p_F\right)^2} \right]
\end{align}
Similarly we have:
\begin{align}
\delta\gamma^- = + \frac{\pi n}{i p_F^2 L^2} \frac{\frac{\pi^2n^2}{p_F^2 L^2} \left(2 - \frac{\pi^2n^2}{p_F^2 L^2} \right)}{\left(m V_A / p_F\right)^2} \frac{1}{1-\exp\left(2ip_F L \sqrt{2 - \frac{\pi^2n^2}{p_F^2 L^2}}\right)} \left[2 \frac{1+(-1)^n \cos\varphi\, \exp\left(ip_F L \sqrt{2 - \frac{\pi^2n^2}{p_F^2 L^2}}\right)}{\frac{\pi n}{p_F L}\sqrt{2 - \frac{\pi^2n^2}{p_F^2 L^2}}} - \frac{1}{\left(m V_A / p_F\right)^2} \right]
\end{align}
Finally, the full solution can be written as 
\begin{align}
\gamma^\pm = \gamma_0^\pm + \delta\gamma^\pm
\end{align}
Note, that $\gamma^\pm$ are complex, and in order to obtain the energies of the bound states we need to take the real parts of those expressions, whereas the imaginary parts yield the broadening. It is also worth mentioning that we are interested in bound states forming close to zero energy, in other words, we consider values of $n \in \mathbb{N}$ such that
$$
\frac{\pi^2n^2}{p_F^2 L^2} \sim 1.
$$
Applying this condition to the corrections obtained above, we get:
\begin{align}
\delta\gamma^\pm \sim \mp \frac{1}{i p_F L} \frac{1}{\left(m V_A / p_F\right)^2} \frac{1}{1-e^{2ip_F L}} \left\{ 2 \left[1+(-1)^n \cos\varphi\, e^{ip_F L} \right] - \frac{1}{\left(m V_A / p_F\right)^2} \right\}
\end{align}
Note that at values $p_F L = \pi q$, with $q \in \mathbb{Z}$ this correction diverges, setting the limits of validity of this perturbative approach. 
\end{document}